# Fast recovery of ion-irradiation-induced defects in $Ge_2Sb_2Te_5$ thin films at room temperature

*Martin Hafermann[1], Robin Schock[1], Chenghao Wan[2,3], Jura Rensberg[1], Mikhail A. Kats[2,3,4], and Carsten Ronning[1]*

[1]*Institute of Solid State Physics, University of Jena, 07743 Jena, Germany*

[2]*Department of Electrical and Computer Engineering,* [3]*Department of Materials Science and Engineering,*

*and* [4]*Department of Physics, University of Wisconsin−Madison, Madison, WI 53706, USA*

## Abstract

Phase-change materials serve a broad field of applications ranging from non-volatile electronic memory to optical data storage by providing reversible, repeatable, and rapid switching between amorphous and crystalline states accompanied by large changes in the electrical and optical properties. Here, we demonstrate how ion irradiation can be used to tailor disorder in initially crystalline $Ge_2Sb_2Te_5$ (GST) thin films via the intentional creation of lattice defects. We found that continuous $Ar^+$-ion irradiation at room temperature of GST films causes complete amorphization of GST when exceeding 0.6 (for rock-salt GST) and 3 (for hexagonal GST) displacements per atom ($n_{dpa}$). While the transition from rock-salt to amorphous GST is caused by progressive amorphization via the accumulation of lattice defects, several transitions occur in hexagonal GST upon ion irradiation. In hexagonal GST, the creation of point defects and small defect clusters leads to disordering of intrinsic vacancy layers (van der Waals gaps) that drives the electronic metal−insulator transition. Increasing disorder then induces a structural transition from hexagonal to rock-salt and then leads to amorphization. Furthermore, we observed different annealing behavior of defects for rock-salt and hexagonal GST. The higher amorphization threshold in hexagonal GST compared to rock-salt GST is caused by an increased defect-annealing rate, i.e., a higher resistance against ion-beam-induced disorder. Moreover, we observed that the recovery of defects in GST is on the time scale of seconds or less at room temperature.

## I. INTRODUCTION

At least theoretically, almost any material can be solidified into at least two distinct states—a disordered amorphous state and an ordered crystalline state. Thus, phase transitions between amorphous and crystalline states may occur for numerous materials. The class of phase-change materials (PCMs), however, is unique in that the phase transition at elevated temperatures can be exceptionally fast, and both phases are (meta)stable at ambient conditions[1]. Furthermore, in most PCMs, the structural transitions are accompanied by electronic transitions, which result in distinct electrical and optical properties between various phases, facilitating their uses in non-volatile electronic memory and optical data storage[1–7]. The most extensively studied PCMs are pseudo-binary $GeTe–Sb_2Te_3$ chalcogenide compounds, particularly $Ge_2Sb_2Te_5$, which provides high switching speed and material stability. $Ge_2Sb_2Te_5$ crystallizes into a metastable rock-salt-like structure and a stable hexagonal structure[8]. The rock-salt structure comprises face-centered cubic (fcc) unit cells, with octahedrally coordinated Te atoms occupying the anion-like fcc sublattice and randomly distributed Ge, Sb, and intrinsic vacancies occupying the cation-like lattice sites[9,10]. Ordering the randomly distributed vacancies (when sufficient energy is provided) drives an insulator-to-metal transition (IMT) caused by the formation of highly ordered vacancy layers within the stable hexagonal phase[11–14]. While the exact atomic arrangement in the hexagonal unit cell is still a matter of debate, cubic close-packed stacking (abcabc) with a nine-plane periodicity comprising pure Te and mixed Ge/Sb basal planes along the c-axis is proposed[11–13,15]. There is a large separation between two adjacent Te layers corroborated by a significantly larger bond length (~ 3.7 °A) compared to the Ge/Sb-Te bond lengths (~ 3 °A). This ordered vacancy layer is accompanied by weak van der Waals forces between low (3) coordinated Te atoms (this is referred to as a van der Waals gap). It is understood that the van der Waals gaps are responsible for metal-like conduction within hexagonal GST (which behaves as a degenerate p-type semiconductor[16]). Furthermore, it has been proposed that the IMT is purely driven by disorder-induced (Anderson) localization[17]. Thus, the electrical and optical properties of GST strongly depend on the degree of disorder, which facilitates tuning these properties by controlling the amount of disorder within the material. Apart from thermal annealing[17], a variety of approaches have emerged for tuning disorder in GST compounds. These include laser excitation[18,19], focused electron beam irradiation[20], tuning of chemical composition[21,22], interface templates[22–24], or the application of pressure[25–30], strain[31,32], and voltage pulses[33–35]. Ion irradiation facilitates precisely tailoring disorder in solids via nuclear collision cascades of the impinging ions; thus, multiple studies have investigated the effect of defect engineering of GST compounds[36–44]. While these studies report on different properties examined with various structural, optical, or electrical techniques either during (*in-situ*) or after (*ex-situ*) ion bombardment, the comparison of the same material property obtained from *in-situ* and *ex-situ* measurements is missing. Performing both *in-situ* and *ex-situ* measurements would yield insights into the annealing behavior of ion-irradiation-induced defects.

Here, we demonstrate a combined approach of *in-situ* optical/electrical and *ex-situ* optical/structural measurements during and after noble-gas ion irradiation of crystalline (rock-salt and hexagonal) GST thin films. We reveal how ion irradiation drives various electronic and structural transitions towards complete amorphization of GST thin films via the intentional creation of structural disorder. Furthermore, our approach allows us to investigate the effect of *in-situ* annealing of defects during ion bombardment. For this, we compare, first, the *in-situ* and *ex-situ* measurements and, second, the influence of the ion flux on the *in-situ* measurements.

## II. ION-BEAM-INDUCED DISORDER: IN-SITU MEASUREMENTS

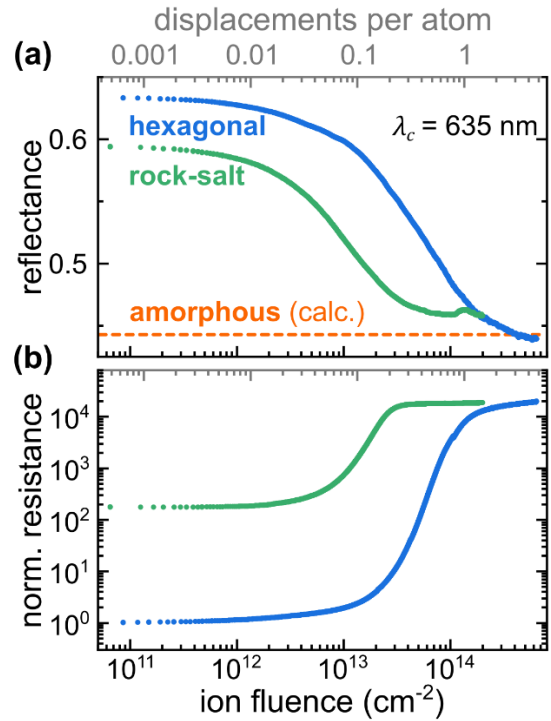

**Figure 1** Simultaneous *in-situ* optical/electrical measurements of ~100-nm-thick GST films on SiO$_2$/Si, initially in either the hexagonal or rock-salt phase: **(a)** Reflectance at $\lambda$ = 635 nm versus ion fluence ($N_i$) of initially hexagonal and rock-salt GST films upon Ar$^+$-ion irradiation at a constant flux of 2 x 10$^{11}$ cm$^{-2}$s$^{-1}$. The dashed orange line corresponds to the calculated reflectance value of a comparable amorphous GST film (using optical data from reference[45]). **(b)** Simultaneously obtained normalized resistance ($R(N_i)/R(0)$) for both phases with the ion fluence dependent resistance $R(N_i)$ and the resistance before irradiation $R(0)$. To account for the different intrinsic resistances of rock-salt and We performed *in-situ* characterizations to investigate the effect of ion irradiation on the electrical and optical properties of GST films that were initially in rock-salt and hexagonal phases, respectively. For this purpose, magnetron-sputtered, annealed GST thin films with a thickness of approximately 100 nm on $SiO_2$/Si substrates were irradiated with Ar$^+$-ions at room temperature while the samples' optical reflectivity and electrical resistance were simultaneously and continuously monitored (for details on the setup see supplementary material). The insulating (~300 nm thick) $SiO_2$ barrier prevents current flow through the substrate. Note that Ar is not chemically incorporated into the GST matrix, which allows the investigation of the effects of induced lattice damage without simultaneous impurity doping. We used an ion energy of 180 keV to create a damage profile that is almost uniform across the film thickness. Note, that we use different ion energies for the specific experiment depending on whether the ions penetrate through the film-substrate-interface (see also supplementary material): While the uniformity of the defect distribution for high ion energies is mandatory for the electrical measurements, maintaining a pristine interface using low ion energies is advantageous for optical spectroscopy measurements. The ion flux was set to $2 \times 10^{11}$ cm$^{-2}$s$^{-1}$, which does not result in substantial thermal heating by the ion beam[46]. In Fig. 1a, we plotted the optical reflectance at $\lambda$ = 635 nm of initially rock-salt and hexagonal GST films upon ion irradiation as a function of ion fluence. To enable comparison with experiments under different conditions (e.g., using different ion energy), we converted the ion fluence into the number of displacements per lattice atom, $n_{dpa}$, (gray upper axis, see supplementary material). Note, however, that $n_{dpa}$ is derived from Monte-Carlo TRIM simulations at absolute zero[47], and defect migration and recombination reduce the residual amount of lattice displacements at elevated temperatures. Since the TRIM simulation returns a depth-dependent value for the number of displacements per incident ion, we use the maximum of this curve to derive $n_{dpa}$. Further, because our *in-situ* experimental setup only detects changes of the optical reflectivity, the absolute reflectance values were calculated for the intrinsic films using the optical constants of the different GST phases[45]. The reflectance of both of the two GST films decreases upon ion bombardment and saturates when $n_{dpa}$ > 0.6 for the rock-salt phase and $n_{dpa}$ > 3 for the hexagonal phase, which agrees well with the amorphization thresholds obtained in similar experiments by Privitera *et al.*[37]. Upon saturation, the reflectance of both films is close to the calculated reflectance of an amorphous (as-deposited) GST film (for details on the calculation see supplementary material).

Simultaneous resistance measurements during ion bombardment are shown in figure 1b. Here, relative changes were compared instead of absolute values due to different contact geometries of each sample. To account for the different intrinsic resistances of rock-salt and hexagonal GST films, the resistance change of the rock-salt sample was normalized to the saturated resistance value of the (amorphized) hexagonal sample, because the specific resistance in the amorphous phase should be similar for both samples. Thus, the resulting resistance difference of intrinsic hexagonal and rock-salt GST is ~200, in good agreement with reported values[17,48]. The resistance increases upon ion irradiation by 4 (2) orders of magnitude and saturates for $n_{dpa}$ values above 1.2 (0.24) for the hexagonal phase (rock-salt phase). These $n_{dpa}$ saturation values are slightly lower than those for the optical data, which might indicate a slightly different impact of the amount of induced disorder in the film to the optical and electrical properties.

## III. TAILORING DISORDER: EX-SITU MEASUREMENTS

We further gain insights into the various structural and electronic transitions in GST that occur upon ion bombardment by performing optical spectroscopy and X-ray diffraction. Here, in contrast to the *in-situ* measurements from above, the measurements were carried out after certain irradiation steps with increasing ion fluence on separate samples. For these *ex-situ* measurements, multiple initially rock-salt and hexagonal GST thin films (d ≈ 100 nm) on Si were irradiated at room temperature with Ar+-ions at an angle of incidence of 7°, with various ion fluences. Optical and structural measurements were performed *ex-situ* on both GST phases minutes to hours after the ion irradiation. Here, we used an ion energy of 55 keV to avoid interface mixing (see supplementary material for more details). Inadvertent interface mixing for higher ion energies may reduce thin-film interference effects, which would hamper the comparability between different samples. Note that each ion fluence (or $n_{dpa}$) corresponds to an individual sample; thus, sample-to-sample variations should be considered.

## A. Disorder-induced progressive amorphization of rock-salt GST

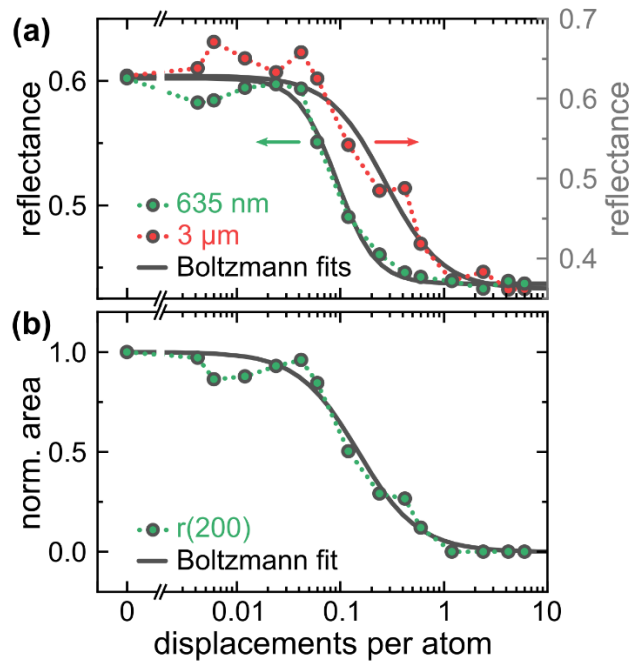

**Figure 2** Ar irradiation of rock-salt GST films: **(a)** Reflectance in the visible (635 nm, green) and infrared (3 μm, red) as a function of displacements per lattice atom $n_{dpa}$, fitted using Boltzmann's sigmoidal equation (solid lines). **(b)** Extracted Gaussian peak area from XRD measurements of the r(200) reflection, normalized to the value of the intrinsic rock-salt GST film. Note, the dotted lines between data points are guides for the eye.

The reflectance change as a function of $n_{dpa}$ at visible (635 nm) and near-IR (3 μm) wavelengths of rock-salt GST films on Si is given in figure 2a. In both cases, the reflectance remains almost unaffected by ion irradiation up to $n_{dpa}$ values of ~0.05. Small deviations can be attributed to sample-to-sample variations. In this low-ion-fluence regime, mainly point defects, such as vacancies and interstitials, are formed, which do not seem to alter the optical properties of rock-salt GST. With further increasing $n_{dpa}$, the reflectance drastically decreases in the visible and saturates at $n_{dpa}$ ~ 0.5, corresponding to complete amorphization of GST. In the near-IR (at 3 μm), however, the effective amorphization threshold shifts to higher $n_{dpa}$ values. Note, that $n_{dpa}$ is used for normalization of the ion fluence and do not resemble different defect concentrations in different depths. A refractive index gradient is present within the GST film due to the non-uniform damage profile of 55 keV Ar⁺-ions (see supplementary material). While visible light probes roughly the first half of the film, at 3 μm the extinction of GST is significantly smaller than in the visible[45]; thus, the reflected light contains information from the entire film and interface. The shift in amorphization threshold is therefore caused by the smaller amount of defects near the film/substrate interface.

We also performed XRD measurements on the same set of samples. Figure 2b shows the peak area of the r(200) reflection, which is indicative of the rock-salt structure, as a function of $n_{dpa}$, where the peak area was extracted using a Gaussian function fitting (for details see supplementary material). Upon ion irradiation, there is a reduction in the intensity of the r(200) peak. Moreover, the inflection point of the r(200) peak is in between the inflection points of reflectance curves at visible and near-IR wavelengths. This corroborates the lower defect concentration near the GST/Si interface due to the non-uniform damage profile, because the XRD pattern also contains information about the entire film. We observed no further peaks in the XRD pattern upon ion irradiation, indicating progressive amorphization of the GST film (see supplementary material). This progressive amorphization likely proceeds by accumulation of point defects and small defect clusters or by direct amorphization of small regions within the GST film.

## B. Disorder-driven electronic and structural transitions in hexagonal GST

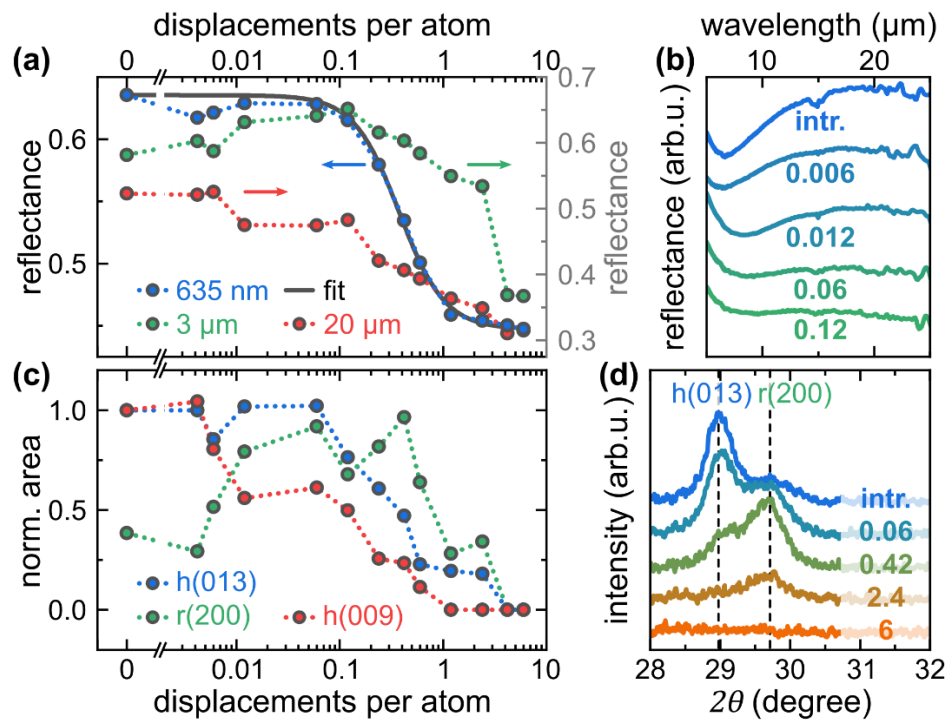

**Figure 3** Ar irradiation of hexagonal GST films: **(a)** Reflectance in the visible (635 nm, blue) and near- (3 μm, green) and mid-infrared (20 μm, red) as a function of displacements per lattice atom, $n_{dpa}$. Only the 635 nm data was fitted using Boltzmann's sigmoidal equation (gray solid line). Note, the dotted lines between data points are guides for the eye. **(b)** Selected reflectance spectra in the infrared before and after irradiation with low $n_{dpa}$ values (blue to green). **(c)** Extracted Gaussian peak areas of some characteristic reflection peaks. The h(009) and h(013) peak areas are normalized to the values corresponding to an intrinsic hexagonal GST film. The r(200) area values are normalized to the value of the intrinsic h(013) reflection to account for different contributions of rock-salt and hexagonal phases. **(d)** XRD patterns of intrinsic (hexagonal) and irradiated GST films with increasing $n_{dpa}$ around the region of the h(013) and r(200) reflections corresponding to the hexagonal and rock-salt phase, respectively.

In figure 3a, we show the reflectance change of hexagonal GST films on Si after Ar⁺-ion irradiation as a function of $n_{dpa}$ at visible (635 nm), near-IR (3 µm), and mid-IR (20 µm) wavelengths. The reflectance in the visible reaches the value of the amorphous phase for $n_{dpa} \approx 1$, which matches well the trend in the *in-situ* measurements (compare figure 1). Due to the non-uniform damage profile within the GST film, the observable amorphization threshold is shifted towards higher $n_{dpa}$ for infrared wavelengths. Figure 3b shows the IR reflectance spectra of hexagonal GST films prior and after irradiation with low $n_{dpa}$ values between 0.006 and 0.12 (blue to green). The optical properties of intrinsic hexagonal GST—a degenerate p-type semiconductor[16]—are dominated by a pronounced Drude contribution in the IR spectral region caused by intrinsic, ordered vacancy layers (van der Waals gaps). Note that the sputtered GST films are polycrystalline; thus, any optical anisotropy is averaged out. The van der Waals gaps are responsible for the high mobility in hexagonal GST due to delocalized states. The Drude contribution results in a high and increasing reflectance with respect to mid-IR wavelengths, which was clearly reduced upon ion irradiation with low $n_{dpa}$ values due to the creation of point defects (vacancies, interstitials, or replacements) and small defect clusters. This decrease of the Drude contribution is caused by a reduction in the free-carrier concentration and/or mobility. Ion irradiation induces a disordering of the inherent vacancy layers[37], which is accompanied by the formation of localized states and a reduction of delocalized states, which leads to a reduction of carrier concentration/mobility and is indeed observed in the infrared spectra for low ion fluences in figure 3b. This disorder-driven metal-to-insulator transition can be regarded as the inverse process of the vacancy-ordering-induced insulator-to-metal transition upon thermal annealing[17].

XRD measurements were performed on the hexagonal GST films after ion irradiation. Figure 3d displays some representative XRD patterns around $2\theta \approx 29°$. Here, only the contributions of the h(013) ($\sim 29.0°$) and the r(200) ($\sim 29.6°$) reflections can be seen. Both reflections can be identified for the intrinsic, annealed GST film; thus, there is a minor portion of rock-salt phase within the intrinsic hexagonal GST film even before ion bombardment. However, the contribution of each reflection changes upon ion bombardment, and thus reveals an increased portion of rock-salt phase compared to hexagonal phase for increasing $n_{dpa}$. There are no reflections for $n_{dpa} > 4$, indicating complete amorphization.

The normalized area of the h(013) reflection, which is given in figure 3c, remains roughly constant upon ion bombardment for $n_{dpa}$ < 0.1 and then gradually decreases until vanishing due to complete amorphization. In contrast, the r(200) reflection area first increases up to $n_{dpa}$ ~ 0.5 before decreasing due to amorphization. This further reveals an increase of the portion of rock-salt phase inside the film for low $n_{dpa}$ values accompanied with a reduction of the hexagonal phase.

Density functional theory (DFT) simulations on vacancy layer formation in disordered crystalline GST[37,49] have shown that the rock-salt structure has a lower formation energy compared to the hexagonal structure when both contain depleted layers with 50% vacancy concentrations. Thus, as the vacancy occupancy of the van der Waals gaps decreases upon ion irradiation to about 50%, the rock-salt structure becomes energetically more favorable. Hence, the energy penalty that occurs upon vacancy disordering in the hexagonal phase is causing the structural transition from hexagonal to rock-salt.

Furthermore, the reduction of the h(009) reflection (figure 3c) is shifted to lower $n_{dpa}$ compared to the h(013) peak, which indicates a loss of the nine-plane periodicity in hexagonal GST[11–13,15,50]. This reveals that the periodicity that is linked to the van der Waals gaps is lost prior to the phase transformation of the hexagonal structure. Moreover, the trend of the h(009) reflection correlates well with the trend for the reflectance at 20 μm (figure 3a), corroborating the link between the van der Waals gaps and the carrier mobility, and thus the Drude response of hexagonal GST.

Note also that the r(200) reflection and the optical reflectance at 3 μm correlate well, because rock-salt GST is more reflective than hexagonal GST[48]. Comparing these trends with that from h(009) and the reflectance at 20 μm suggests that there is no direct correlation between the metal-insulator transition and the structural hexagonal to rock-salt transition, which is in good agreement with previous studies[17,37].

## IV. IN-SITU ANNEALING OF DEFECTS DURING ION IRRADIATION

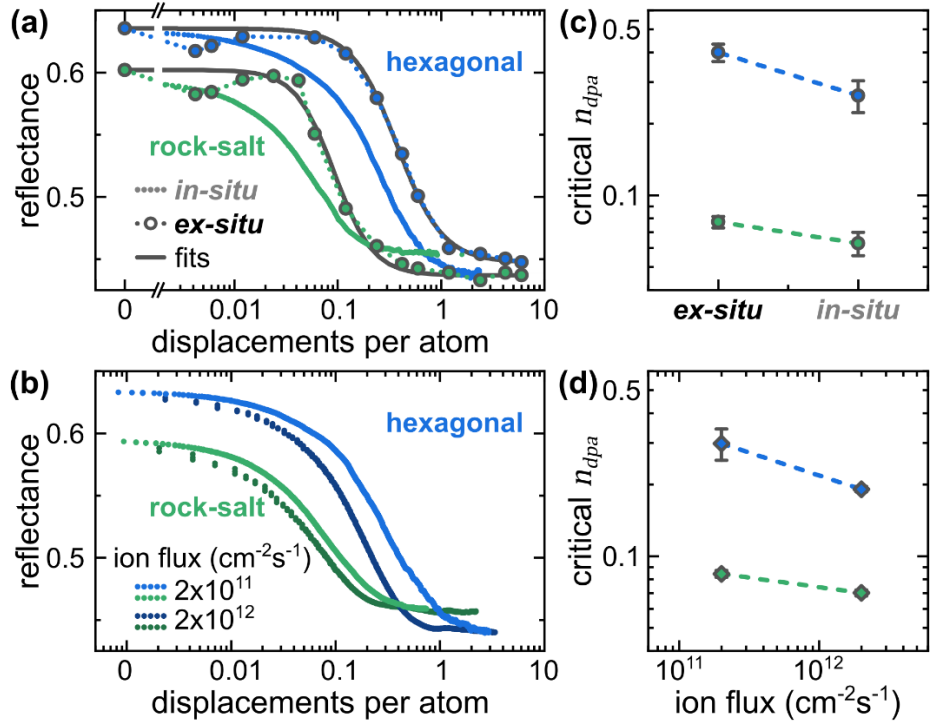

**Figure 4 (a)** Comparison of the reflectance at 635 nm of initially rock-salt (green) and hexagonal (blue) GST films on Si upon Ar irradiation for *in-situ* (constant flux of $2 \times 10^{11}$ cm$^{-2}$s$^{-1}$, 100 keV, 45°) and *ex-situ* experiments (55 keV, 7°) as a function of displacements per lattice atom. The *ex-situ* measurements were fitted using Boltzmann's sigmoidal equation (gray lines). **(b)** *In-situ* optical measurements of initially rock-salt (green) and hexagonal (blue) GST films on Si upon Ar irradiation (180 keV, 45°) for different ion fluxes. **(c,d)** Extracted critical number of displacements per atom $n_{dpa}$ (defined as inflection point of reflectance curves) to compare (c) *in-situ* and *ex-situ* experiments and (d) different ion fluxes.

Because all measurements were carried out at room temperature, the induced lattice damage may partially recover due to thermal recombination of mobile defects. Thus, the amount of created residual disorder is a consequence of two opposing effects: the rate of atomic displacements upon ion irradiation versus the rate of defect migration and recombination at a given temperature. The 635 nm reflectance of ~100 nm crystalline GST films on Si upon Ar$^+$-ion irradiation is given in figure 4a for *in-situ* and *ex-situ* measurements, respectively. The basic difference between the in-situ and ex-situ measurements is the timeframe between irradiation and optical measurement. While the impact on the reflectance is measured directly during ion bombardment in the *in-situ* measurements, there are typically several minutes to hours between irradiation and *ex-situ* optical measurements. Note, that further measurements after weeks/months do not show further changes. For ease of comparison, the ion energy was chosen so that the irradiation defect depth profiles are similar for *ex-situ* (ion energy: 55 keV, angle of incidence: 7°) and *in-situ* measurements (ion energy: 100 keV, angle of incidence: 45°, see also supplementary material). Note that the *in-situ* measurements shown in figure 4 are from different samples than the measurements shown in figure 1.

The reduction in reflectance is shifted to higher $n_{dpa}$ in the *ex-situ* measurements for both crystalline GST phases. This can be attributed to annealing of lattice defects through migration and recombination at room temperature. Furthermore, for *ex-situ* measurements, the reflectance is roughly constant up to $n_{dpa} \approx 0.05$ and $\approx 0.1$ for the rock-salt and hexagonal phase, respectively. Thus, point defects and small defect clusters might be able to recover almost completely at room temperature within the ~hours between irradiation and optical measurement. Conversely, there is a gradual reduction of the *in-situ* reflectance for both phases at low ion fluences. In the *in-situ* measurements, apparently, the time between subsequent ion impacts into the same volume is not sufficient to recover even for small amounts of point defects before the ongoing ion bombardment leads to the formation of larger, more stable defects.

Furthermore, *in-situ* optical measurements were performed with different ion fluxes (ion energy: 180 keV, angle of incidence: 45°) of initially rock-salt and hexagonal GST films on Si (figure 4b), which corresponds to an altered rate of atomic displacements. However, the rate of defect migration and recombination should be independent of the ion flux and only depend on the sample's temperature. Note that for the ion fluxes used here, significant macroscopic heating of the sample caused by the ion beam can be neglected; however, local heating by the impinging ions may occur on the microscopic scale and influences the annealing effect. Apparently, increasing the ion flux by an order of magnitude leads to a shift of the amorphization threshold to smaller $n_{dpa}$ values for both phases resulting from an increased rate of atomic displacements. Moreover, this shift is smaller for the rock-salt than for the hexagonal phase, indicating a lower annealing rate. This is further corroborated by an overall lower amorphization threshold of rock-salt GST. By neglecting directional effects of ion-irradiated polycrystalline films, the rate of atomic displacements should be independent of the crystal structure of chemically identical media. However, the mobility and recombination of defects strongly depends on the crystal structure of GST. This is a consequence of different diffusion paths of defects and dislocated lattice atoms that clearly depend on the local atomic arrangement. It was reported that the van der Waals gaps in hexagonal GST serve as preferential sinks where displaced atoms are able to recombine[39]. Due to unfavorable Te-Te antibonds and low mobility of Te atoms, primarily Ge and Sb tend to migrate and recombine in the gaps. As these gaps are not found in rock-salt GST, the resulting annealing rate is higher in the hexagonal phase, which leads to an increased critical $n_{dpa}$ for amorphization.

We extracted critical values of $n_{dpa}$ , which is defined as the inflection point of the reflectance curves, from Lorentzian fits to the derivatives of both *ex-situ* and *in-situ* measurements and the different ion fluxes (see supplementary material), which are depicted in figure 4c and d, respectively. While *ex-situ* measurements result in a ~1.5 times higher critical $n_{dpa}$ for hexagonal GST compared to *in-situ*, this ratio is reduced to ~1.2 for the rock-salt phase. Coincidentally, in the *in-situ* measurements very similar values were obtained by comparing the different ion fluxes, thus, different times between subsequent ion impacts into the same volume. These observations further corroborate a higher defect mobility and recombination rate of hexagonal GST compared to rock-salt GST. The comparison of *ex-situ* and *in-situ* measurements suggests that the annealing of defects at room temperature is on the time scale of seconds or less.

## V. CONCLUSION

In summary, we have demonstrated that ion irradiation can be utilized to tailor disorder in a phase-change material. *In-situ* optical and electrical measurements during noble gas ion bombardment of initially rock-salt and hexagonal GST thin films revealed complete amorphization at room temperature accompanied by large changes of the reflectance and resistance. *Ex-situ* optical and structural measurements revealed several structural and electronic transitions in GST thin films. While the transition from rock-salt to amorphous GST is caused by progressive amorphization via the accumulation of lattice defects, several transitions occur in hexagonal GST upon ion irradiation. The creation of point defects and small defect clusters leads to disordering of the inherent vacancy layers (van der Waals gaps) that are responsible for the metal-like conduction inside hexagonal GST. The vacancy disordering process drives the metal-insulator transition via the formation of localized states and a reduction of carrier mobility. XRD measurements revealed that further ion irradiation leads to a transition from the hexagonal phase to the rock-salt phase, until eventually amorphization is reached at high ion fluences. By comparing *in-situ* and *ex-situ* measurements, and *in-situ* measurements with different ion fluxes, we found that hexagonal GST has a higher amorphization threshold compared to rock-salt GST, which is indicative of a higher resistance against ion-beam-induced disorder and thus a more stable hexagonal phase.

## VI. EXPERIMENTAL DETAILS

### A. Sample Preparation

GST thin films were grown by DC magnetron sputtering on (100)-silicon substrates and thermally oxidized $SiO_2$/Si substrates. The magnetron sputtering conditions were 300 V and ~ 50 mA with an argon pressure of $5 \times 10^{-3}$ mbar at room temperature. Crystallization to the rock-salt phase of the as-deposited, amorphous GST films was achieved by heating the samples (heating rate: 10 K/min) up to 250°C in an Ar atmosphere while the amplitude of the reflection at 635 nm was monitored. Heating the samples up to 350°C leads to the transition into hexagonal GST.

### B. Methods

Crystalline GST films were irradiated with $Ar^+$-ions using a 400 kV single-ended ion accelerator with ion energies from 55 to 180 keV and beam fluxes from $10^{11}$ to $10^{12}$ $cm^{-2}s^{-1}$. The reflectivity and resistivity of the samples in the *in-situ* measurements were simultaneously monitored during ion bombardment. The light intensity reflected by the sample from a red laser diode (635 nm) is detected via a biased Si detector. Changes to the resistivity are measured in a two-probe method using contacts made of conductive lacquer and a Keithley 2000 multimeter.

Reflectance spectra after ion irradiation were (*ex-situ*) obtained by a combination of UV-Vis spectroscopy using a Varian Cary 5000 UV-Vis-NIR spectrometer and FTIR spectroscopy using a Varian 640-IR spectrometer. Structural characterization of the irradiated GST films was carried out using a Bruker D8 Discovery X-ray diffractometer in θ – 2θ mode and a Cu-Kα micro X-ray source.

### Supplementary Material

See the supplementary material for the *in-situ* experimental setup, monte-carlo TRIM calculations, reflectance calculations, reflectance spectra, XRD pattern, ion energy dependence, and determination of critical $n_{dpa}$.

## Author Contributions

M.H., R.S., C.W., and J.R. performed the measurements. The manuscript was written by M.H. with feedback from all authors. All authors have given approval to the final version of the manuscript.

## Funding.

This work was financed by the Deutsche Forschungsgemeinschaft (DFG) through grant Ro1198/21-1 and by a collaborative exchange program of the Deutscher Akademischer Austauschdienst (DAAD) through grant 57386606. M. A. K. acknowledges support from the Office of Naval Research (N00014-20-1-2297) and Northrop Grumman Corporation, Space Systems, NG Next Basic Research.

## Disclosures.

The authors declare no conflicts of interest.

## Data availability.

Data underlying the results presented in this paper are not publicly available at this time but may be obtained from the authors upon reasonable request.

# SUPPORTING INFORMATION TO

# Fast recovery of ion-irradiation-induced defects in Ge2Sb2Te5 thin films at room temperature

*Martin Hafermann[1], Robin Schock[1], Chenghao Wan[2,3], Jura Rensberg[1], Mikhail A. Kats[2,3,4], and Carsten Ronning[1]*

[1]*Institute of Solid State Physics, University of Jena, 07743 Jena, Germany*

[2]*Department of Electrical and Computer Engineering,* [3]*Department of Materials Science and Engineering,*

*and* [4]*Department of Physics, University of Wisconsin–Madison, Madison, WI 53706, USA*

## S-I. IN-SITU EXPERIMENTAL SETUP

The experimental setup, which is schematically depicted in figure S1, allows to simultaneously monitor changes in the reflectivity and/or resistivity of a sample during ion bombardment. Optical measurements were conducted with a red laser diode ($\lambda_c$ = 635 nm). The incoming and outgoing laser beams are guided through a window (near-normal incidence) while the sample is tilted to 45° with respect to the ion beam inside the irradiation vacuum chamber. In this way, changes to the reflectivity of the sample upon ion irradiation are directly observed by detecting the light intensity reflected by the sample via a Si detector. Simultaneously, changes to the resistivity are monitored by contacting the sample via conductive lacquer to a Keithley 2000 multimeter. Note that no predefined contacts were used; thus, absolute resistance values cannot be compared due to different contact geometries between several samples. For these *in-situ* experiments, the beam current was measured via Faraday cups with a defined area to calculate the ion fluence.

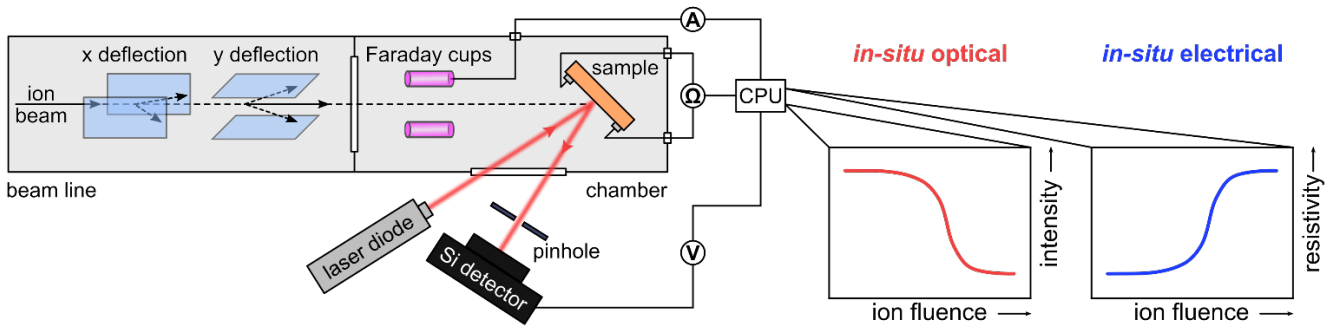

**Figure S1** Schematic of the experimental setup used for in-situ optical/electrical measurements during ion irradiation with respective symbolic curves.

## S-II. MONTE-CARLO TRIM SIMULATIONS

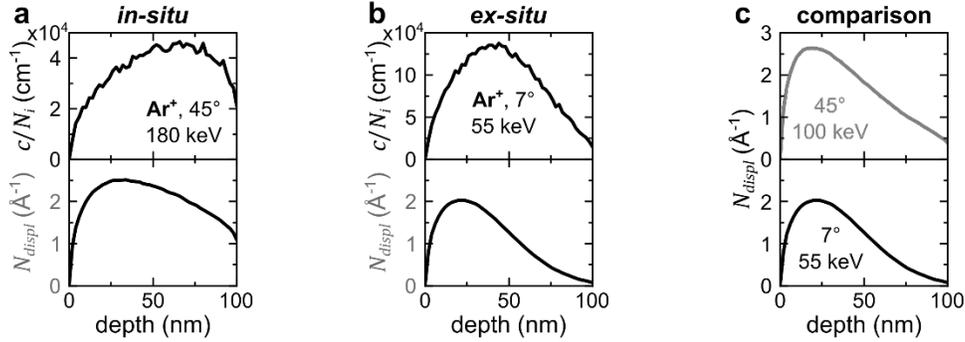

**Figure S2 (a,b)** Ar concentration normalized to ion fluence $c/N_i$ (top) and total displacements per incident ion $N_{displ}$ (bottom) as a function of depth as simulated with TRIM[1] for (a) 180 keV Ar⁺-ions at 45° incidence angle used in the *in-situ* measurements and (b) 55 keV Ar⁺-ions at 7° incidence angle used in the *ex-situ* measurements. **(c)** Total displacements per incident ion $N_{displ}$ as a function of depth for 100 keV Ar⁺-ions at 45° incidence (top) and 55 keV Ar⁺-ions at 7° incidence angle (bottom). For the comparison of *in-situ* and *ex-situ* measurements, the energy for the *in-situ* measurement was set to 100 keV to achieve a very comparable damage distribution (compared to the 55 keV irradiation), because the angle of incidence changes for the different setups.

To account for different $N_{displ}$ of the respective measurements, the number of displacements per lattice atom was calculated with:

$$n_{dpa} = \frac{N_{displ}^{max} \times N_I}{n_{at}}$$

using the maximum of the number of displacements per incident ion $N_{displ}^{max}$ (with ion fluence $N_I$). Note that the atomic density $n_{at}$ slightly changes upon irradiation. As an approximation, for the calculation of $n_{dpa}$ only the initial values of the respective GST phases were assumed.

## S-III. ION ENERGY DEPENDENCE OF LATTICE DEFECTS

Multiple GST films ($d \approx 100$ nm) on SiO₂/Si and Si were irradiated with Ar⁺-ions at various energies. While *in-situ* optical measurements were performed during ion bombardment of GST films on Si, the electrical measurements required the SiO₂/Si substrate. The results from both measurement setups are depicted in figure S3. First, *in-situ* optical measurements were performed at 45° incidence angle for beam energies of 100 keV and 180 keV at GST thin films on Si. The latter energy was used to ensure nearly uniform damage distribution across the entire film. Consequently, an inevitable amount of Ar⁺-ions reaches the substrate and

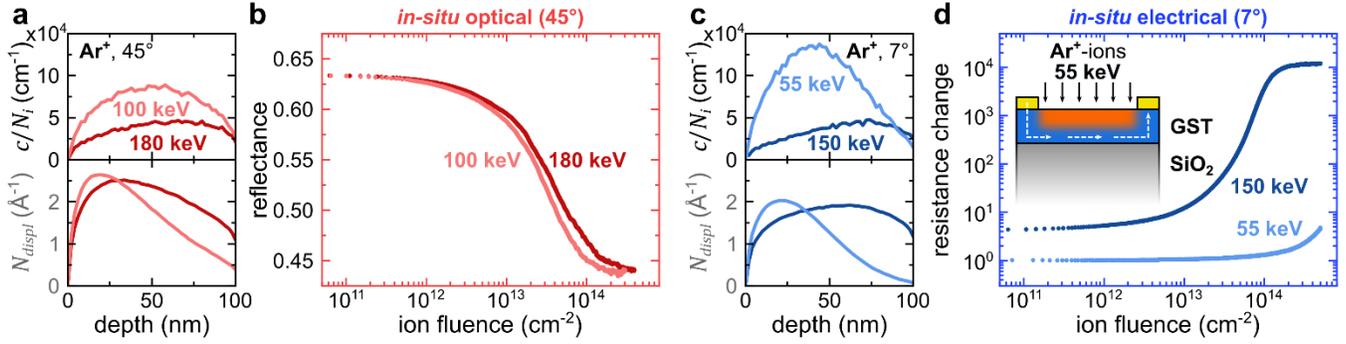

**Figure S3 (a)** Ar concentration per ion fluence $c/N_i$ (top) and total displacements per incident ion $N_{displ}$ (bottom) for Ar$^+$-ions into hexagonal GST with ion energies 100 keV and 180 keV and 45° incidence angle. **(b)** Reflectance at 635 nm of initially hexagonal GST films on Si upon Ar irradiation at a constant flux of $2 \times 10^{11}$ cm$^{-2}$s$^{-1}$ for the ion energies and incidence angle of (a). **(c)** Ar concentration per ion fluence $c/N_i$ (top) and total displacements per incident ion $N_{displ}$ (bottom) as a function of depth for Ar$^+$-ions into hexagonal GST with ion energies 55 keV and 150 keV and 7° incidence angle. **(d)** Resistance change of a GST film on SiO$_2$/Si at a constant flux of $2 \times 10^{11}$ cm$^{-2}$s$^{-1}$, which was first irradiated with an Ar ion energy of 55 keV and further irradiated with 150 keV at an incidence angle of 7°.

possible interface mixing might influence the GST layer. Reducing the beam energy to 100 keV and keeping the angle at 45° leads to a non-uniform damage profile while most Ar$^+$-ions remain in the film and interface mixing can be neglected. The reflectance change upon ion irradiation for these energies is given in figure S3b. The reflectance decreases down to the expected amorphous value in both cases with only minor differences in the critical amount of induced lattice defects. This can be attributed to the different number of total displacements per ion for both energies (compare figure S3a). As the red laser light ($\lambda$ = 635 nm) only probes roughly the first 50 nm of the GST film (with the extinction coefficient being ~ 4 for both crystalline phases), the reflectance signal contains mainly information about the first half of the film. Thus, the slightly higher amount of displacements near the surface for the 100 keV irradiation compared to the higher beam energy of 180 keV explains the slightly smaller critical ion fluence for amorphization.

A similar examination was performed for *in-situ* electrical measurements on approximately 100 nm thick GST films on SiO$_2$/Si. As these measurements were not performed simultaneously to the *in-situ* optical examination, here, an incidence angle of 7° was chosen. As the depth distribution of damage caused by impinging ions is also determined by their angle of incidence, the beam energies were adjusted properly. Similar damage profiles can be achieved by using 55 keV and 150 keV ions (see figure S3c). The resistance change of a GST film upon 55 keV ion irradiation is depicted in figure S3d. Even for comparable high ion fluences well above previously obtained amorphization thresholds the resistance only slightly increases. On the other hand, subsequent irradiation with 150 keV ions of the same sample leads to a drastic increase in resistance to the expected change caused by amorphization. As the induced lattice damage at the interface of the GST and SiO$_2$ films is negligible for a beam energy of 55 keV, a non-damaged path with low resistance remains that mainly affects the total sheet resistance (see inset of figure S3d). Therefore, the beam energy and thus the penetration depth plays an important role for investigating disorder in GST films via electrical methods but has no significant impact on the *in-situ* optical measurements at 635 nm for rather thick films ($\gtrsim$ 100 nm). However, below the optical gap of GST, i.e., its absorption-free thus transmissive spectral region (mainly in the IR), inadvertent interface mixing for higher ion energies may reduce thin-film interference effects that are then independent of changes in the refractive indices. Therefore, 55 keV Ar$^+$-ions were used for the *ex-situ* measurements to avoid interface mixing.

## S-IV. CALCULATION OF REFLECTANCE

Light propagation through single and multilayer thin films can be described by using the Fresnel equations. Here, we used a MATLAB based code that relies on the transfer matrix method to calculate reflectance values from the respective refractive indices of the individual layers. Considering Fresnel and Maxwell's equations, a simple 2x2 matrix can be used to describe the electric field within a single layer. The description of multi-layers is then the product of each layer's matrix and can be solved while also considering internal reflections and interference of coherent light waves. Refractive index values were taken from references[2–4].

## S-V. REFLECTANCE SPECTRA

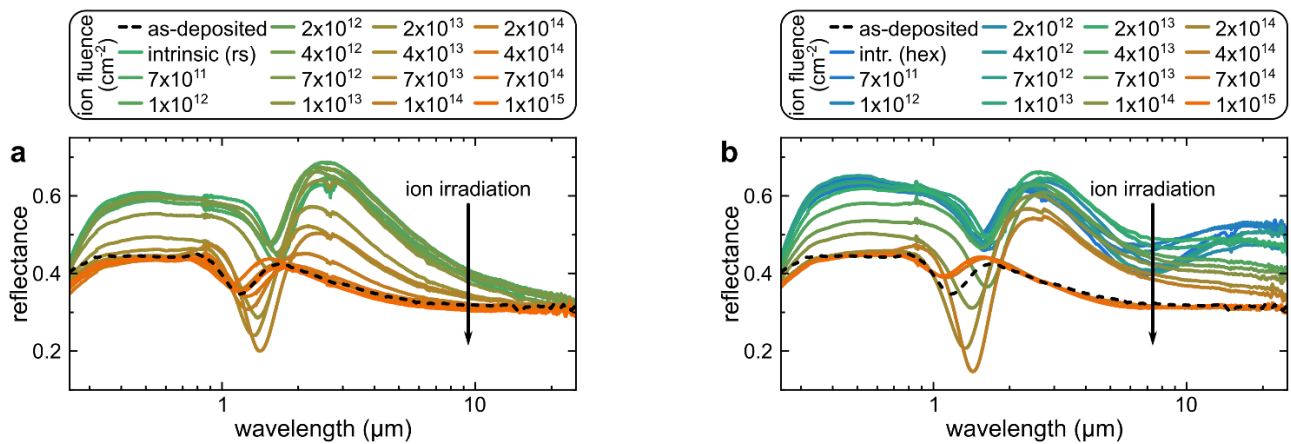

**Figure S4** Reflectance spectra of **(a)** initially rock-salt and **(b)** initially hexagonal GST thin films (d ≈ 100 nm) on Si before and after Ar⁺ ion irradiation with various, increasing ion fluences. Additionally, a reflectance spectrum of a comparable as-deposited, amorphous GST thin film is given (black dashed line).

## S-VI. X-RAY DIFFRACTION PATTERN

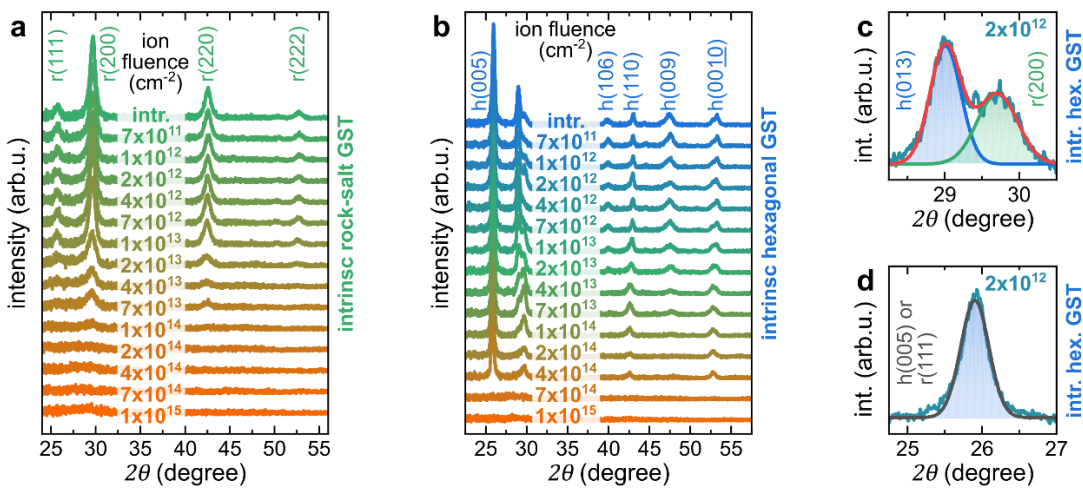

**Figure S5** XRD patterns of **(a)** initially rock-salt and **(b)** initially hexagonal GST thin films (d ≈ 100 nm) on Si before and after Ar⁺ ion irradiation with various, increasing ion fluences. **(c,d)** Exemplary multiple Gaussian peak fits of (c) the h(013) (blue) and r(200) (green) reflections and (d) the superposition of the h(005) and r(111) reflection of the initially hexagonal GST film irradiated with $2 \times 10^{12}$ cm⁻², respectively.

## S-VII. COMPARISON OF DIFFERENT EXPERIMENTS

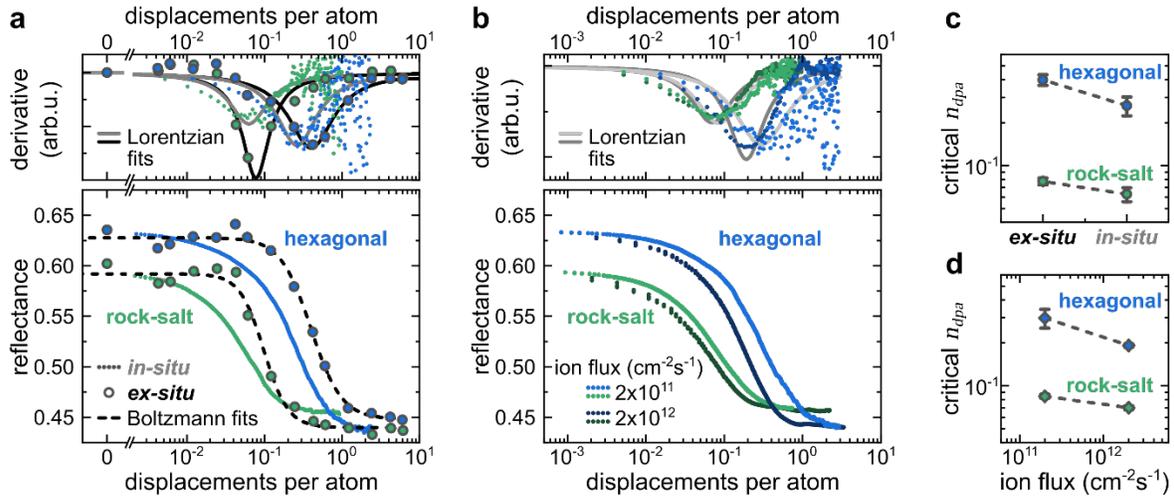

**Figure S6 (a)** Comparison of the reflectance at 635 nm of initially rock-salt (green) and hexagonal (blue) GST films on Si upon Ar irradiation for *in-situ* (constant flux of 2 ×10¹¹ cm⁻²s⁻¹, 100 keV, 45°) and *ex-situ* experiments (55 keV, 7°) as a function of displacements per lattice atom. *Ex-situ* measurements were fitted using Boltzmann's sigmoidal equation (black dashed lines). Additionally, respective derivatives of all curves are fitted (on a logarithmic scale) with Lorentzian distribution functions (solid lines). **(b)** *In-situ* optical measurements of initially rock-salt (green) and hexagonal (blue) GST films on Si upon Ar irradiation (180 keV, 45°) for different ion fluxes together with respective derivatives and Lorentzian fits. **(c,d)** Extracted critical number of displacements per atom $n_{dpa}$ corresponding to the peak position of the fitted Lorentzian functions to compare (c) *in-situ* and *ex-situ* experiments and (d) different ion fluxes.